\UseRawInputEncoding
\documentclass[%
 reprint,
 twocolumn,
nofootinbib,
 amsfonts,amsmath,amssymb,
 aps,
prd,
floatfix,
showpacs,
longbibliography
]{revtex4-2}

\usepackage{graphicx}
\usepackage{dcolumn}
\usepackage{bm}
\usepackage{wasysym}
\usepackage{soul}
\usepackage{MnSymbol}
\usepackage{xfrac}
\usepackage[dvipsnames]{xcolor}
\definecolor{ReflexBlue}{rgb}{ .0902,.0902,.5882}
\usepackage{hyperref}
\hypersetup{
    colorlinks=true,
    linkcolor=ReflexBlue,
    filecolor=magenta,      
    urlcolor=ReflexBlue,
    citecolor=ReflexBlue,
}
\begin{document}


\title{Renormalization of $\langle\phi^2\rangle$ at the inner horizon of rotating, accreting black holes}

\author{Tyler McMaken}
 \email{Tyler.McMaken@colorado.edu}
\author{Andrew J. S. Hamilton}
 \email{Andrew.Hamilton@colorado.edu}
\affiliation{%
 JILA and Department of Physics, University of Colorado, Boulder, Colorado 80309, USA
}%
\date{\today}

\begin{abstract}
Classically, the inner horizon of a perturbed, rotating black hole undergoes an instability known as mass inflation, wherein the spacetime curvature diverges as a result of hyper-relativistic crossing streams of ingoing and outgoing radiation. The generic outcome of this instability is currently believed to be a strong, spacelike singularity, potentially alongside a weak, null singularity surviving at late times. However, the quantum back-reaction in this regime has yet to be fully calculated for a realistic black hole spacetime. Here we consider a massless quantized scalar field $\phi$ over the inflationary Kasner spacetime, a recently developed model for the inner horizon geometry of a rotating, accreting black hole. With this spacetime, we use numerical adiabatic regularization to calculate $\langle\phi^2\rangle_\text{ren}$, the renormalized coincidence limit of the two-point correlation function, as a pointer to the behavior of the quantum stress-energy tensor. $\langle\phi^2\rangle_\text{ren}$ is generically found to be nonzero near the inner horizon, divergent where the curvature classically diverges, and larger for smaller black hole spins or accretion rates.
\end{abstract}

\maketitle

\section{\label{sec:int}Introduction}

What happens inside the event horizon of a black hole? If the classical laws of general relativity are to be believed, then the answer is well known. For vacuum models of black holes that possess a charge (Reissner-Nordstr{\"o}m) or angular momentum (Kerr), below the event horizon lies a second horizon, called the \emph{inner} or \emph{Cauchy horizon}, which marks the boundary of causality. Below this inner horizon, the vacuum models predict a wormhole and a naked timelike singularity, but since the 1960s, it has been predicted that when the effects of matter and radiation are included in these models, a singularity will form at the inner horizon, precluding any possibility of causality violation \cite{pen68,sim73}.

The reason for a classical singularity at the inner horizon of a black hole stems from the fact that this horizon is a surface of infinite blueshift\textemdash an infalling observer at an inner horizon would see the entire history or future of the Universe flash before their eyes as the energy of incoming radiation becomes classically unbounded. The result is a phenomenon known as \emph{mass inflation}, an inner horizon instability marked by the divergence of the spacetime curvature and the quasi-local internal mass parameter. The instability was first shown to be an inevitable result of nonlinear perturbations by Poisson and Israel in 1990 \cite{poi90,bar90}, and since that time, later models have been developed to confirm and generalize the result \cite{ori91,ori92,gne93,bra95,ori96,ori98,ori99,bur02,bur03,daf05,ham10,ham11a,ham11b,ham11c,ham17,che19,mcm21}.

The main discrepancy that still exists within the classical picture of mass inflation concerns the choice of initial conditions, which dictates the nature of the singularity formed at the inner horizon. If the black hole forms from an eternally isolated gravitational collapse, the result is a weak, null singularity \cite{ori91,ori92,bra95,ori96,ori98,ori99,daf05}. In contrast, if any small, continuous accretion of matter or radiation is included, the resulting singularity is strong and spacelike \cite{gne93,bur02,bur03,ham10,ham11a,ham11b,ham11c,ham17,che19,mcm21}. In reality, it may be the case that both these singularities are present in different sectors of the interior. But regardless, these pictures of the inner horizon must be incomplete, since they do not take into account any quantum effects.

As a quantized field theory, gravity is non-renormalizable, since Feynman diagram calculations at all loop orders produce divergences that only get worse the higher one goes in the perturbative expansion. However, as an effective field theory, one-loop divergences can be absorbed by renormalization of the next-order parameters, so that quantum gravitational effects can be calculated provided the characteristic frequencies of the gravitational background do not exceed the Planck frequency \cite{bir82}. The one-loop approach is usually given the name ``semiclassical gravity,'' wherein matter fields are quantized while the background spacetime is treated classically. Such an approach has led to celebrated predictions like the existence of Hawking and Unruh radiation.

For calculations of quantum field theories (QFTs) in curved spacetimes, the quantity of greatest interest is the renormalized expectation value of the field's stress-energy tensor $\langle T_{\mu\nu}\rangle$, since it contributes to a back-reaction to the spacetime geometry via the semiclassical field equations
\begin{equation}\label{eq:semiclassicalEinstein}
    G_{\mu\nu}=8\pi\left(T_{\mu\nu}^{\text{classical}}+\langle T_{\mu\nu}\rangle\right).
\end{equation}
The squared amplitude $\langle\phi^2\rangle$ of vacuum fluctuations is often used as a proxy for the calculation of $\langle T_{\mu\nu}\rangle$, since $\langle\phi^2\rangle$ contains fewer divergences that need to be renormalized yet still provides meaningful information about vacuum polarization effects. In particular, $\langle\phi^2\rangle$ can be used to the determine the trace of $\langle T_{\mu\nu}\rangle$ for conformally coupled scalar fields, and $\langle\phi^2\rangle$ additionally provides insight about spontaneous symmetry breaking in a given background spacetime \cite{and89}.

In the context of semiclassical black hole interiors, most works have considered only two-dimensional or charged black holes. In the two-dimensional case, the quantum back-reaction at the inner horizon leads to a divergence of the stress-energy and the formation of a spacelike singularity \cite{fla97,fro05}. In the charged, spherically symmetric case, early work indicated that $\langle T_{\mu\nu}\rangle$ would also likely diverge at the inner horizon to produce a spacelike singularity, but could also remain regular in certain cases \cite{bir78,bal93}. In the past few years, an explosion of works studying semiclassical Reissner-Nordstr{\"o}m mass inflation have shown that the ingoing null component $\langle T_{vv}\rangle$ yields a non-zero value at the inner horizon, and switching to a time coordinate which is regular through the inner horizon shows that the renormalized stress-energy tensor does physically diverge there \cite{lan19,tay20,zil20,hol20,bar21,zil21,kle21}. Back-reaction from $\langle T_{vv}\rangle$ alone is generally believed to cause a strong curvature singularity \cite{hol20}, though there is no complete semiclassically consistent solution to verify this yet. What has been shown is that a dynamical semiclassical inner horizon will evaporate outwards over time (along with a slower, inward-evaporating outer horizon), leading to a self-consistent steady state that inevitably contains a curvature singularity \cite{bar21,arr21a,arr21b}.

For the case of spinning black holes, far fewer works have been published analyzing quantum effects at the inner horizon. In the simplified vacuum case (the Kerr metric), before mass inflation was even understood classically, Hiscock argued from symmetry and conservation conditions that $\langle T_{\mu\nu}\rangle$ must diverge on either the ingoing or outgoing inner horizon, a result confirmed in the decades following \cite{his80,ott00}. More recently, $\langle T_{\mu\nu}\rangle$ was calculated for the inner horizon of rotating ($2+1$)-dimensional black holes, which was found to result in a spacelike singularity \cite{cas19}. Finally, a new set of works by \citeauthor{zil22a} have shown that for a ($3+1$)-dimensional Kerr black hole, the flux components of $\langle T_{\mu\nu}\rangle$ do generically diverge at the inner horizon, with the specific choice of spin and polar angle determining whether the null flux components of $\langle T_{\mu\nu}\rangle$ are positive or negative.

Despite the success of the aforementioned studies in calculating quantum effects in Kerr black holes, the Kerr metric is not a realistic model when describing the asymptotic regime near the inner horizon of an astrophysical black hole, which in general is modified by the perturbative effects of accretion. Instead of applying a quantum field over some vacuum spacetime solution, we here consider the inflationary Kasner spacetime, a recently developed model for the inner horizon behavior of a rotating black hole with accretion \cite{mcm21}. The details and assumptions of this model are described in Sec.~\ref{subsec:spamet}. Then, we proceed to calculate the renormalized vacuum polarization ${\langle\phi^2\rangle_\text{ren}}$. The quantization procedure is detailed in Sec.~\ref{subsec:quafiemod}, while the renormalization procedure is detailed in Sec.~\ref{sec:renpro}. Since the wave equation for this model cannot be solved analytically, a numerical framework is used, as described in Sec.~\ref{sec:numres}. This framework is first applied to a simplified, isotropic case in Sec.~\ref{subsec:flrwren}, after which the results for the inner horizon are presented in Sec.~\ref{subsec:infkasren}. The paper then concludes with a discussion of the results of the calculation of ${\langle\phi^2\rangle_\text{ren}}$ in Sec.~\ref{sec:dis}, leaving the calculation of ${\langle T_{\mu\nu}\rangle_\text{ren}}$ to future work.

\section{\label{sec:proset}Problem Setup}

\subsection{\label{subsec:spamet}Spacetime metric}

The spacetime geometry near the inner horizon of a rotating, accreting black hole can be modeled using a homogeneous metric\footnote{Throughout this paper we use the ${({-}{+}{+}{+})}$ metric signature and geometric units where ${c=G=\hbar=M_\bullet=1}$.} \cite{mcm21}:
\begin{equation}\label{eq:metric}
    ds^2=-a_0^2(t)dt^2+\sum_{i=1}^3a_i^2(t)(dx^i)^2,
\end{equation}
where the time-dependent scale factors satisfy
\begin{equation}\label{eq:scalefactors}
    a_0^2=c_1t\ \text{e}^{t^2},\qquad a_1^2=c_1t^{-1}\text{e}^{t^2},\qquad a_2^2=a_3^2=t^2,
\end{equation}
for positive time $t$ and positive constant $c_1$. This metric is called the \emph{inflationary Kasner} metric because of its asymptotic resemblance to the well-known Kasner metric first proposed one hundred years earlier \cite{kas21}. In particular, for $t$ above unity, the metric approximates a Kasner metric with Kasner exponents ${(p_1,p_2,p_3)}={(1,0,0)}$, and as $t$ decreases on its way down to the spacelike singularity at ${t=0}$, the spacetime ``bounces'' to a Kasner metric with Kasner exponents ${(-\sfrac{1}{3},\ \sfrac{2}{3},\ \sfrac{2}{3})}$.

The coordinates and constants used in the line element of Eqs.~(\ref{eq:metric})--(\ref{eq:scalefactors}) are chosen for their convenience for the QFT calculations done here. They are related to those of Ref.~\cite{mcm21} by:
\begin{equation}\label{eq:tc1}
    t=T^{1/2},\qquad c_1=\frac{1}{4\pi\Phi_0T_0^{1/2}\text{e}^{T_0}},
\end{equation}
along with a suitable rescaling of the spatial coordinates. The astrophysical properties of the black hole (viz., the spin $a$, the inner horizon Boyer-Lindquist radius $r_-\equiv{1-\sqrt{1-a^2}}$, and the initial accretion rate $u$) are related to these constants via:\footnote{The initial energy density $\varPhi_0$ depends more generally on the observer's polar coordinate $\theta$ and the difference in the speed of ingoing and outgoing streams \cite{ham11b}, but the inclusion of these parameters provides no more precision than the inflationary Kasner approximation already affords.}
\begin{equation}\label{eq:T0Phi0}
    T_0=\frac{r_-^3-3r_-^2+a^2r_-+a^2}{u(r_-^2+a^2)^2},\qquad\varPhi_0\approx\frac{u^2T_0}{4\pi}.
\end{equation}

To comment briefly on the physical interpretation of this metric, note that the inflationary Kasner model holds only in the regime asymptotically close to the inner horizon of a rotating, accreting black hole. Above this horizon, spacetime is well-approximated by the Kerr metric. But once an observer approaches the inner horizon, they will experience the mass inflation phenomenon described in Sec.~\ref{sec:int}, corresponding to the radial collapse of the inflationary Kasner metric as $t$ decreases from its initial value of ${t_0\equiv\sqrt{T_0}}$ (which is generally quite large, since $T_0$ is inversely proportional to the generally tiny accretion rate $u$, in units where the black hole mass is 1) until reaching ${t\sim\sqrt{\sfrac{1}{2}}}$. Then, the inflationary Kasner model predicts a bounce in the spacetime, wherein the radial collapse reverses and the collapse in the isotropic directions proceeds towards a strong, spacelike singularity at ${t=0}$.

Several key assumptions underlie the inflationary Kasner model; in particular, the near-inner-horizon spacetime is assumed to be homogeneous and sourced by the self-similar, asymptotically small accretion of a collisionless, null fluid. First, the assumption of homogeneity comes about from the fact that during mass inflation, the curvature and stress-energy exponentiate rapidly while the Boyer-Lindquist coordinates $r$ and $\theta$ remain frozen at their inner horizon values. One should not imagine that a vast, structureless swath of homogeneous spacetime lies hidden within accreting black holes; rather, homogeneity applies locally for each near-inner-horizon observer as the entire structure of the inflationary Kasner model passes by within a fraction of a second of their proper time. Second, the assumptions about accretion are a direct result of the process of mass inflation, which accelerates any infalling matter to relativistic velocities along the radial direction, rendering any other contributions to Einstein's equations irrelevant until ${t/t_0}$ reaches below $\sim\sqrt{10^{-5}}$ \cite{mcm21}. Finally, the assumption of a tiny accretion rate should hold for any black hole at late times (indeed, even the cosmic microwave background would source such accretion).

\subsection{\label{subsec:quafiemod}Quantum field modes}

Let $\phi(x)$ be a canonically quantized, neutral scalar field, evaluated at a spacetime point $x$, that satisfies the Klein-Gordon wave equation \cite{bir82}
\begin{equation}\label{eq:waveeq_phi}
    \left(\mbox{\large$\square$}-m^2-\xi R\right)\phi=0,
\end{equation}
where ${\mbox{\large$\square$}\equiv g^{\mu\nu}\nabla_\mu\nabla_\nu}$ is the d'Alembert operator, $m$ is the mass of the field quanta, $\xi$ is a numerical factor indicating the strength of the coupling between the scalar field and the gravitational field, and $R$ is the Ricci scalar curvature. For the inflationary Kasner metric, ${R(x)=0}$, so the calculation of ${\langle\phi^2\rangle_{\text{ren}}}$ here will not depend on $\xi$. Later, the massless case will be assumed, but in order for the adiabatic expansion of Sec.~\ref{subsec:adireg} to be valid, $m$ must not be set to 0 until the end of the calculation\footnote{Numerically, a variety of values for the mass were tested until convergence in the ${m\to0}$ limit was achieved; for the present analysis, ${m=10^{-4}}$ was found to be more than sufficient.}.

The homogeneity of the inflationary Kasner spacetime motivates a decomposition of the quantum field operator $\phi$ into a set of modes indexed by wavevectors ${\bm{k}\in\mathbb{R}^3}$:
\begin{equation}\label{eq:phi_modedecomp}
    \phi=\frac{1}{(2\pi)^{3/2}}\int\!d^3\bm{k}\left[A_{\bm{k}}\psi_{\bm{k}}(t)\ \text{e}^{i\bm{k}\cdot\bm{x}}+A_{\bm{k}}^\dagger\psi_{\bm{k}}^*(t)\ \text{e}^{-i\bm{k}\cdot\bm{x}}\right].
\end{equation}
The creation and annihilation operators $A_{\bm{k}}^\dagger$ and $A_{\bm{k}}$ will then satisfy the usual commutation relations
\begin{equation}
    [A_{\bm{k}},A_{\bm{k'}}]=[A_{\bm{k}}^\dagger,A_{\bm{k'}}^\dagger]=0,\qquad[A_{\bm{k}},A_{\bm{k'}}^\dagger]=\delta^3(\bm{k}-\bm{k'}),
\end{equation}
provided the mode functions $\psi_{\bm{k}}$ satisfy the Wronskian condition
\begin{equation}\label{eq:wronskian}
    \dot{\psi}^*_{\bm{k}}\psi_{\bm{k}}-\psi^*_{\bm{k}}\dot{\psi}_{\bm{k}}=i\frac{a_0}{a_1a_2a_3}=it^{-1}.
\end{equation}
Here and throughout, an overdot represents differentiation with respect to the coordinate time $t$. Introducing the auxiliary function
\begin{equation}
    f_{\bm{k}}(t)\equiv\left(\frac{a_1a_2a_3}{a_0}\right)^{1/2}\psi_{\bm{k}}(t)=t^{1/2}\psi_{\bm{k}}(t),
\end{equation}
Eq.~(\ref{eq:waveeq_phi}) leads to the dynamical equation
\begin{equation}\label{eq:waveeq_ft}
    \ddot{f}_{\bm{k}}+\left(\omega_{\bm{k}}^2+\sigma\right)f_{\bm{k}}=0,
\end{equation}
where the frequency $\omega_{\bm{k}}(t)$ and $\bm{k}$-independent geometrical background term $\sigma(t)$ are defined by
\begin{align}\label{eq:omega}
    \omega_{\bm{k}}^2(t)&\equiv a_0^2\left(\sum_{i=1}^{3}\frac{k_i^2}{a_i^2}+m^2\right),\\
    \sigma(t)&\equiv\frac{1}{4}\left[\frac{2\ddot{a}_0}{a_0}-\frac{3\dot{a}_0^2}{a_0^2}+\sum_{i=1}^{3}\frac{\dot{a}_i^2}{a_i^2}+a_0^2\left(4\xi-1\right)R\right].\label{eq:sigma}
\end{align}
In the isotropic limit, $\omega_{\bm{k}}$ reduces to the standard definition of frequency used for Friedmann-Lema\^{i}tre-Robertson-Walker (FLRW) universes \cite{par74}. With the scale factors of Eq.~(\ref{eq:scalefactors}), the anisotropic background term $\sigma(t)$ simplifies to $1/(4t^2)$, and the wave Eq.~(\ref{eq:waveeq_ft}) has no known solution in terms of analytic functions. Therefore, the mode functions $f_{\bm{k}}$ must be found by numerically solving the wave equation for each choice of wavevector $\bm{k}$. The boundary conditions are set by the choice of vacuum state, the discussion of which is deferred to Sec.~\ref{subsec:chovacsta}.

\subsection{\label{subsec:asy}Asymptotics}

While the inflationary Kasner wave equation has no general closed-form solution, here we comment briefly on three relevant asymptotic regimes: the adiabatic regime (${t/t_0\gg1}$), where the spacetime behaves adiabatically but is too far from the inner horizon singularity to match the behavior of a black hole; the initial inflationary regime ${t/t_0\to1}$, where both the inflationary Kasner and Kerr metrics are valid just above the inner horizon; and the final collapse regime (${t/t_0\to0}$), where the spacetime behavior is dominated by the collapse of the metric toward the spacelike singularity.

\subsubsection{\label{subsubsec:adi}Adiabatic regime}

When ${t/t_0}$ exceeds unity, the exponential terms in the inflationary Kasner scale factors of Eq.~(\ref{eq:scalefactors}) will dominate, provided ${t_0\gg1}$ (as is true for any astrophysical black hole below the Thorne limit \cite{tho74}, since $t_0$ scales as ${u^{-1/2}}$ for the generally tiny accretion rate ${u\ll1}$). The resulting metric, with an exponential term over the temporal and radial sectors, has the same Rindler-type form of Lass's radar coordinates for Minkowski space (when $x$ and ${T\equiv t^2}$ are swapped) \cite{las63}. As a Kasner universe, the metric asymptotically approaches the well-studied case of Kasner exponents ${(1,\ 0,\ 0)}$. But even more simply, the coordinate transformation
\begin{align}
    \tilde{T}&\equiv\sqrt{\frac{c_1}{t}}\text{ e}^{\frac{t^2}{2}}\sinh(x),\quad&&\tilde{X}\equiv\sqrt{\frac{c_1}{t}}\text{ e}^{\frac{t^2}{2}}\cosh(x),\nonumber\\
    \tilde{Y}&\equiv t\ y,\quad&&\tilde{Z}\equiv t\ z
\end{align}
will bring the metric to the form
\begin{equation}
    -d\tilde{T}^2+d\tilde{X}^2+d\tilde{Y}^2+d\tilde{Z}^2.
\end{equation}
in the large-$t$ limit. Thus, the standard QFT approach to flat spacetime applies in this limit, provided the appropriate vacuum state is supplied.

\subsubsection{\label{subsubsec:ini}Initial inflationary regime}

In the massless limit, as ${t/t_0}$ approaches unity from below (in fact, for any value of $t$ in the range ${\text{e}^{-t_0^2/3}\lesssim t\lesssim t_0}$, provided $t_0\gg1$), the frequency function in the wave equation will be dominated by the radial $k_1$ term, which is quadratic in $t$. The general solution to the wave equation in terms of the mode functions ${\psi_{\bm{k}}=f_{\bm{k}}t^{-1/2}}$ can be written in terms of zeroth-order Bessel functions:
\begin{equation}
    \psi_{\bm{k}}=A_{\bm{k}}J_0\left(\frac12k_1t^2\right)+B_{\bm{k}}Y_0\left(\frac12k_1t^2\right),
\end{equation}
for complex coefficients $A_{\bm{k}}$ and $B_{\bm{k}}$.

\subsubsection{Final collapse regime}
As ${t/t_0}$ approaches zero, the exponential terms in the inflationary Kasner scale factors of Eq.~(\ref{eq:scalefactors}) become negligible, which results in a standard Kasner metric with normalized Kasner exponents ${(-\sfrac{1}{3},\ \sfrac{2}{3},\ \sfrac{2}{3})}$. Asymptotically, the wave Eq.~(\ref{eq:waveeq_ft}) simplifies to
\begin{equation}\label{eq:waveeq_ft_collapse}
    \ddot{f}_{\bm{k}}+\left(\frac{c_1k_{\perp}^2}{t}+\frac{1}{4t^2}\right)f_{\bm{k}}=0,
\end{equation}
with a general solution to the mode functions ${\psi_{\bm{k}}=f_{\bm{k}}t^{-1/2}}$ given by zeroth-order Bessel functions with argument $\sqrt{4k_{\perp}^2c_1t}$. But since ${t_0\gg1}$, the constant $c_1$ and therefore the subleading $t^{-1}$ term from Eq.~(\ref{eq:waveeq_ft_collapse}) is exponentially suppressed by the factor $\text{e}^{-t_0^2}$, so that the mode solutions further reduce to
\begin{equation}
    \psi_{\bm{k}}(t)=A_{\bm{k}}+B_{\bm{k}}\ln(t),
\end{equation}
for complex coefficients $A_{\bm{k}}$ and $B_{\bm{k}}$.

\section{\label{sec:renpro}Renormalization Procedure}

The quantity of interest is the probability density of vacuum fluctuations, given by vacuum expectation value of the squared field operator, denoted $\langle0|\phi^2(x)|0\rangle$ (or more concisely, $\langle\phi^2\rangle$). Formally, this quantity can be defined as the coincidence limit of a suitable two-point correlation function,
\begin{equation}\label{eq:phisq_def}
    \langle\phi^2\rangle\equiv\frac{1}{2}\lim_{x'\to x}G^{(1)}(x,x'),
\end{equation}
where
\begin{equation}
    G^{(1)}(x,x')\equiv\langle0|\{\phi(x)\phi(x')\}|0\rangle
\end{equation}
is the Hadamard Green function, defined with anticommutator brackets $\{\}$. In the case of the mode expansion described in Sec.~\ref{subsec:quafiemod}, the field variance can be na\"ively calculated as
\begin{equation}\label{eq:phisq_bare}
    \langle\phi^2\rangle_{\text{bare}}=\frac{1}{(2\pi)^{3}t}\int\!d^3\bm{k}\ |f_{\bm{k}}(t)|^2.
\end{equation}
However, the bare integral in Eq.~(\ref{eq:phisq_bare}) is quadratically divergent. In order to obtain a physical, finite result, some renormalization procedure must be employed, as described in the next subsection.

Since the mode solutions to Eq.~(\ref{eq:waveeq_ft}) cannot be expressed in an analytic form and must instead be solved numerically, most of the standard analytic renormalization techniques (such as dimensional regularization) cannot be used. One robust technique for curved spacetimes, known as point-splitting, has recently been implemented numerically by Levi and Ori under the name \emph{pragmatic mode-sum regularization} (PMR) \cite{lev15,lev16}. Such a technique requires only that the background admits some symmetry (Killing field) to permit a mode expansion, and the present case of homogeneous symmetry (translational splitting) has been carried out successfully for the FLRW metric \cite{bel21}. However, the $x$-splitting variant of PMR works well only for isotropic backgrounds\textemdash for the inflationary Kasner metric, the presence of two independent scale factors renders the PMR method ineffective or perhaps even impossible.\footnote{In particular, whereas the generalized transform ${[\mathcal{T}G^{(1)}_{\text{DS}}(\varepsilon)](k)}$ in the case of isotropic $x$-splitting has kernel ${\text{sinc}(k\varepsilon)}$ and can be written explicitly, cylindrical $x$-splitting requires at most two transforms, with kernels proportional to ${\cos(k\varepsilon)}$ and ${J_0(k\varepsilon)}$, and no generalized Hankel transforms have been found for the divergent pieces $\varepsilon^{-2}$ and ${\ln(\varepsilon)}$ that do not also diverge in $k$-space.} Instead, we use the technique known as adiabatic regularization, as described below.

\subsection{\label{subsec:adireg}Adiabatic regularization}

Instead of renormalizing $\langle\phi^2\rangle$ as a whole or at the level of the two-point function, consider what happens if the mode functions themselves are renormalized before any mode integrals are performed. In particular, the goal will be to find a Wentzel-Kramers-Brillouin-type (WKB-type) expansion of the mode solutions ${f_{\bm{k}}(t)}$. These solutions can then be subtracted off from the numerically-obtained solutions ${f_{\bm{k}}(t)}$ to yield a finite integral by construction. This procedure, known as adiabatic regularization, was developed by Parker and Fulling in the 1970s and has been found to provide a consistent means of renormalization, particularly in the case of homogeneous spacetimes \cite{par74,ful74a,ful74b}.

Although adiabatic regularization lacks a manifestly covariant formulation, it is expected to be robust for spacetimes with a high degree of symmetry (such as the present case with homogeneity and 2D isotropy), and it benefits from the simplicity of its computations and the intuitive clarity of its physical interpretation. Further, the scheme has been shown to be equivalent to the DeWitt-Schwinger point-splitting method for massive scalar fields on any Bianchi Type I spacetime \cite{mat18}. Nonetheless, the results presented here still may be subjected to a degree of scrutiny, since they have not been directly compared against an axiomatic, covariant construction \cite{wal77}. However, the main ambiguity in the adiabatic scheme arises from the choice of the leading-order frequency of Eq~(\ref{eq:zerothorder}) below, and such a choice mainly corresponds to the well-known mass scale ambiguity for massless scalar fields, which should not affect the sought-after temporal dependence of ${\langle\phi^2\rangle}$ \cite{bel20}.

Eq.~(\ref{eq:waveeq_ft}) possesses the formal WKB-type solutions
\begin{equation}\label{eq:fk_WKB}
    f_{\bm{k}}(t)=\frac{\exp\left[-i\int^t\!dt'\ W_{\bm{k}}(t')\right]}{\sqrt{2W_{\bm{k}}(t)}},
\end{equation}
where the WKB approximate frequency $W_{\bm{k}}(t)$ satisfies the nonlinear equation
\begin{equation}\label{eq:Wk}
    W_{\bm{k}}^2=\omega_{\bm{k}}^2+\sigma-\frac12\left(\frac{\ddot{W}_{\bm{k}}}{W_{\bm{k}}}-\frac32\frac{\dot{W}_{\bm{k}}^2}{W_{\bm{k}}^2}\right).
\end{equation}
Note that $f_{\bm{k}}$ contains an arbitrary phase factor associated with the lower bound of the integral in Eq.~(\ref{eq:fk_WKB}). At this stage, the shift from $f_{\bm{k}}$ to $W_{\bm{k}}$ is nothing more than a change of variables; the WKB-type form of Eq.~(\ref{eq:fk_WKB}) has the distinct advantage that the Wronskian condition of Eq.~(\ref{eq:wronskian}) is automatically satisfied if $W_{\bm{k}}$ is chosen to be real and nonnegative.

Under the adiabatic approximation, if the spacetime is slowly varying, any derivative terms in Eq.~(\ref{eq:Wk}) will be small compared to the squared frequency $\omega_{\bm{k}}^2$, so a zeroth-order approximation is to substitute
\begin{equation}\label{eq:zerothorder}
    W_{\bm{k}}^{(0)}\equiv\omega_{\bm{k}}
\end{equation}
[note that the background term ${\sigma(t)}$, Eq.~(\ref{eq:sigma}), has adiabatic order 2 (as defined below) and therefore vanishes in the zeroth-order limit along with the explicit derivative terms on the right-hand side of Eq.~(\ref{eq:Wk})]. Higher-order solutions may then be derived by iteration. The next-highest order reads:
\begin{equation}\label{eq:Wk(2)}
    W_{\bm{k}}^{(2)}=\sqrt{\omega_{\bm{k}}^2+\sigma-\frac12\left(\frac{\ddot{\omega}_{\bm{k}}}{\omega_{\bm{k}}}-\frac32\frac{\dot{\omega}_{\bm{k}}^2}{\omega_{\bm{k}}^2}\right)}.
\end{equation}
The superscript $(A)$ attached to the WKB approximate frequency $W_{\bm{k}}^{(A)}$ denotes the adiabatic order $A$ of the function. The $A^{\text{th}}$ adiabatic order is defined by considering the replacement ${t\to \epsilon t}$ (where the adiabatic parameter $\epsilon$ will be taken to 1 at the end of the calculation) and performing an expansion in powers of $\epsilon$ to obtain terms up to order $\epsilon^A$. Practically, terms of $A^{\text{th}}$ adiabatic order are those with up to $A$ time derivatives of the metric.

The key feature of adiabatic regularization is that in the adiabatic limit ${\epsilon\to0}$ (or equivalently, ${\bm{k}\to\infty}$), the adiabatic expansion of the mode solutions to the wave equation should match the exact mode solutions. Since this limit is precisely the regime where $\langle\phi^2\rangle$ contains ultraviolet divergences, subtracting the adiabatic term involving $W_{\bm{k}}^{(A)}$ from the term involving the exact solutions $W_{\bm{k}}$ should yield a finite, renormalized result that can be integrated.

More precisely, the renormalized field variance is
\begin{equation}
    \langle\phi^2\rangle_\text{ren}=\frac{1}{(2\pi)^3t}\int\!d^3\bm{k}\left[\frac{1}{2W_{\bm{k}}}-\left(\frac{1}{2W_{\bm{k}}}\right)^{(A)}\right].
\end{equation}
According to the standard prescription for $\langle\phi^2\rangle$ renormalization \cite{bir82}, only terms up to adiabatic order 2 (viz., all orders containing terms that yield divergent integrals) need to be subtracted for the solution to be consistent with the results obtained from renormalization of the bare constants in the Lagrangian. Utilizing Eq.~(\ref{eq:Wk(2)}), the result is:
\begin{equation}\label{eq:fksq(2)}
    \left(\frac{1}{2W_{\bm{k}}}\right)^{(2)}=\frac{1}{2\omega_{\bm{k}}}-\frac{\sigma}{4\omega_{\bm{k}}^3}+\frac{\ddot{\omega}_{\bm{k}}}{8\omega_{\bm{k}}^4}-\frac{3\dot{\omega}_{\bm{k}}^2}{16\omega_{\bm{k}}^5}.
\end{equation}
By construction, the terms in Eq.~(\ref{eq:fksq(2)}) that would diverge when integrated over $\bm{k}$ exactly cancel the divergences from the exact mode solutions $1/(2W_{\bm{k}})$.

Additionally, since the integrand is even in $k_1$ and isotropic in the $k_2$-$k_3$ plane, the integral simplifies in cylindrical coordinates (with ${k_{\perp}\equiv\sqrt{k_2^2+k_3^2}}$) to 
\begin{widetext}
\begin{equation}\label{eq:phisq_ren}
    \langle\phi^2\rangle_\text{ren}=\lim_{\Lambda\to\infty}\left(\frac{1}{2\pi^2t}\int_0^\Lambda\!\int_0^\Lambda\!dk_1dk_{\perp}k_{\perp}\left[\frac{1}{2W_{\bm{k}}}-\frac{1}{2\omega_{\bm{k}}}+\frac{\sigma}{4\omega_{\bm{k}}^3}-\frac{\ddot{\omega}_{\bm{k}}}{8\omega_{\bm{k}}^4}+\frac{3\dot{\omega}_{\bm{k}}^2}{16\omega_{\bm{k}}^5}\right]\right).
\end{equation}
\end{widetext}
Eq.~(\ref{eq:phisq_ren}) is the expression used in Sec.~\ref{sec:numres} to calculate the vacuum polarization effects in the inflationary Kasner spacetime, first by numerically solving for $W_{\bm{k}}$ via Eq.~(\ref{eq:Wk}) (with initial conditions provided in the next subsection) and then integrating up to a suitably large choice for the ultraviolet cutoff parameter $\Lambda$.

\subsection{\label{subsec:chovacsta}Choice of vacuum state}

For any curved spacetime, the concept of ``particles'' will not necessarily hold the same meaning for different observers. Thus, any calculation in this framework must make the observer-dependent choice of what defines the vacuum state. Such a choice is equivalent to specifying boundary conditions for the wave equation on the spacetime.

Traditionally, the vacuum state for a stationary black hole spacetime is defined by imposing an initial condition to the wave equation along the spacetime's past null boundaries, where one can naturally specify free wave solutions with respect to some affine parameter along those boundaries. Physically, one can then connect the vacuum state to the standard Minkowski vacuum seen by stationary observers at infinity. However, in the present case, the mode expansion for the inflationary Kasner spacetime remains valid only for observers arbitrarily close to the inner horizon. The solution to the inflationary Kasner wave equation as $t$ approaches infinity has no physical meaning, not only because the spacetime is not stationary, but also because once $t$ becomes larger than $t_0$, the metric must be replaced with the Kerr metric if one wishes to describe an astrophysical black hole.

In the absence of a clear natural choice of vacuum state within the inflationary Kasner spacetime, three options present themselves as physically viable choices:
\begin{enumerate}
    \item The \emph{adiabatic vacuum} \cite{par69,bir82} defined at some time $t_A$ consists of purely positive frequency modes with respect to an adiabatic mode expansion [Eq.~(\ref{eq:fk_WKB})], given $t_A$ lies in a regime where the spacetime is slowly varying.
    \item A \emph{Minkowski vacuum} can be defined via the asymptotic behavior of the inflationary Kasner metric for ${t/t_0\gg1}$ (see Sec.~\ref{subsubsec:adi}), provided the appropriate coordinate transformation and mode decomposition that would allow for the mixing of $\text{e}^{i\bm{k}\cdot\bm{x}}$ and $\text{e}^{-i\bm{k}\cdot\bm{x}}$ waves.
    \item The \emph{Unruh state} \cite{unr76}, which reproduces the predictions of Hawking radiation for stationary black holes \cite{can84,bal84,bal01}, can be defined at the Kerr past null boundaries, propagated through the spacetime until it reaches asymptotically close to the inner horizon, then matched onto the inflationary Kasner spacetime via a suitable coordinate transformation and mode decomposition.
\end{enumerate}

Since all three of these vacuum states have transparent, physical interpretations within the black hole spacetime, they all should lead to roughly similar vacuum expectation values, at least when considering the temporal dependence of the effects of particle production from the rapid evolution of the spacetime curvature during mass inflation and collapse. The vacuum state specifically tailored to study the production of physical particles (in the sense of the experiences of a comoving particle detector in a dynamic, homogeneous spacetime) is the adiabatic vacuum state of Option 1, and it is this state that will be used in the present analysis. The most physically authentic choice for the vacuum state would likely be Option 3, especially considering its recent success in the calculation of the renormalized stress-energy tensor at the Kerr inner horizon \cite{zil22a,zil22b}. However, the transition from Unruh modes to an equivalent set of $\bm{k}$-modes in the inflationary Kasner regime is complex and nontrivial, and the authors are currently working on methods to apply the states of Options 2 and 3 in a follow-up work.

The adiabatic family of vacuum states $|0^{(A)}\rangle$, which forms the focus of the present analysis, is defined at $A^{\text{th}}$ adiabatic order such that the annihilation operator $A_{\bm{k}}^{(A)}$ satisfies
\begin{equation}
    A_{\bm{k}}^{(A)}|0^{(A)}\rangle=0
\end{equation}
and designates an exact mode decomposition of the field operator $\phi$ akin to Eq.~(\ref{eq:phi_modedecomp}) \cite{par74}. The adiabatic terminology comes into play because the quantized mode solutions to the wave equation are matched to an adiabatic expansion of those modes (to order $A$) at a time ${t=t_A}$. It should be noted that despite the terminology, the adiabatic state is not merely an approximate vacuum; on the contrary, it represents an exact solution to the wave equation, with the well-defined choice of positive frequency modes motivated by the state one would get from a given adiabatic expansion.

The adiabatic vacuum has several distinct advantages as a physical vacuum state. First, it only requires a matching at a specific time $t_A$, which can be taken in this case to be far from the inflationary Kasner bounce so that the effects of mode distortion from the changing spacetime are minimized in the construction of the quantized field modes. Second, and more importantly, in the adiabatic limit ${\epsilon\to0}$, a comoving particle detector in this vacuum will detect a spectrum that falls off faster than any inverse power of the momentum $\bm{k}$ \cite{par69,par74}. Since particle number is an adiabatic invariant, as long as
\begin{equation}\label{eq:adiabatic_condition}
    \omega_{\bm{k}}\gg\frac{\dot{\omega}_{\bm{k}}}{\omega_{\bm{k}}},
\end{equation}
the excitation of the large-$\bm{k}$ modes will be highly suppressed, and the adiabatic vacuum will exactly match the physical vacuum definition of particles. But even for non-adiabatic portions of a spacetime, the minimization postulate encoded by the statements above implies that the adiabatic definition of creation operators will approximately match that of physical particles throughout the evolution of the spacetime, up to adiabatic order $A$. For the inflationary Kasner spacetime, the adiabatic condition~(\ref{eq:adiabatic_condition}) holds in the strict sense for large $k_1$, $k_2$, or $k_3$, and it holds more generally when ${t\ll1}$ or ${t\gg1}$. For large $t$, since ${\omega_{\bm{k}}\sim\text{e}^{t^2/2}}$ and ${\dot{\omega}_{\bm{k}}/\omega_{\bm{k}}\sim t}$, the adiabatic condition is met even for times as small as ${t\sim5}$, where the frequency $\omega_{\bm{k}}$ exceeds its logarithmic derivative already by several orders of magnitude.

\begin{figure}[t!]
    \includegraphics[width=0.95\columnwidth]{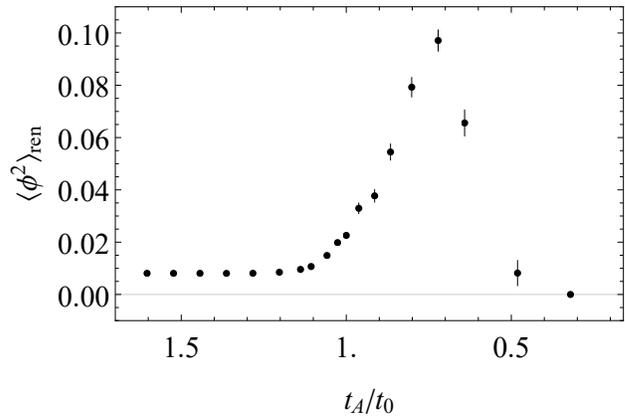}
    \caption{The renormalized variance ${\langle\phi^2(t)\rangle_{\text{ren}}}$ evaluated at the inflationary Kasner time ${t=1\approx0.32t_0}$ for various choices of the adiabatic matching time $t_A$. The constants used are the black hole spin ${a=0.96}$ and initial accretion rate ${u=0.02}$, for which ${t_0\approx3.1}$.}
  \label{fig:phisq_ren_tA}
\end{figure}

To demonstrate the robustness of the adiabatic state used here, Fig.~\ref{fig:phisq_ren_tA} shows the computed value of ${\langle\phi^2(t)\rangle_{\text{ren}}}$ at a certain intermediate time (${t/t_0\approx0.32}$) for a family of different adiabatic vacuum states parametrized by the adiabatic matching time $t_A$. In this plot, the renormalized field variance ${\langle\phi^2\rangle_{\text{ren}}}$ is found to be exactly 0 when the adiabatic vacuum time $t_A$ is the same as the evaluated time $t$ (the rightmost point on the plot), since the field is in vacuum by definition. But if the adiabatic vacuum state is chosen to begin at a time $t_A$ earlier than the point being evaluated, ${\langle\phi^2\rangle_{\text{ren}}}$ obtains a nonzero value corresponding to the polarization of the vacuum accomplished by the spacetime's evolution from $t_A$ to $t$. For a choice of $t_A$ far enough into the past, the value of ${\langle\phi^2\rangle_{\text{ren}}}$ asymptotes to a constant, indicating that the chosen vacuum begins in a suitably adiabatic regime. Note that for the choice of constants used throughout this paper, $t_0$ (the starting time for the mass inflation epoch and the point of matching between the Kerr and inflationary Kasner metrics) takes on a value of about 3.1, which is not quite large enough to preside in the adiabatic regime. The vacuum time $t_A$ must therefore be chosen to be distinct from (and farther in the past than) $t_0$ (in particular, ${t_A=5\approx1.6t_0}$), though it is still close enough to $t_0$ that the inflationary Kasner model should still hold reasonably well.

In order to perform the renormalization calculations in the adiabatic vacuum state, the wave equation [in this case, Eq.~(\ref{eq:Wk})] is solved using a standard numerical integrator, with the Cauchy initial conditions
\begin{subequations}\label{eq:inits}
\begin{equation}
    \frac{1}{2W_{\bm{k}}(t_0)}=\left(\frac{1}{2W_{\bm{k}}(t_0)}\right)^{(A)},
\end{equation}
\begin{equation}
    \partial_t\left(\frac{1}{2W_{\bm{k}}(t_0)}\right)=\partial_t\left(\frac{1}{2W_{\bm{k}}(t_0)}\right)^{(A)},
\end{equation}
\end{subequations}
where the quantities with superscript $(A)$ are truncated at adiabatic order ${A=2}$ by Eq.~(\ref{eq:Wk(2)}) to ensure consistency with the renormalization scheme. As mentioned in Sec.~\ref{subsec:adireg}, the mode functions ${W_{\bm{k}}(t)}$ are used for numerical calculations instead of ${f_{\bm{k}}(t)}$ because the Wronskian condition of Eq.~(\ref{eq:wronskian}) is guaranteed to be satisfied as long as ${W_{\bm{k}}(t)}$ is constrained to be real and nonnegative.

\section{\label{sec:numres}Numerical Results}

To demonstrate how the present numerical framework of adiabatic regularization can be applied to a spacetime (and to confirm its validity), the technique is first carried out for a simplified yet comparable Bianchi Type I spacetime. The chosen spacetime (FLRW model with scale factor $t^{1/3}$) has the advantage that the renormalized variance is known analytically from several independent renormalization techniques and can also be calculated using a numerical adiabatic regularization scheme identical to the one presented here. The numerical results for this simplified case are shown in Sec.~\ref{subsec:flrwren} to be entirely consistent with the analytic solutions. Then, the main results of the renormalization for the inflationary Kasner spacetime are presented in Sec.~\ref{subsec:infkasren}.

\subsection{\label{subsec:flrwren}FLRW renormalization}

The Friedmann-Lema\^{i}tre-Robertson-Walker (FLRW) metric is a special case of the homogeneous metric of Eq.~(\ref{eq:metric}) where the scale factors are completely isotropic. For the present case, assume the scale factors
\begin{equation}\label{eq:flrw}
    a_0=1,\ a_1=a_2=a_3=t^{1/3},
\end{equation}
which correspond to a flat FLRW universe with a classical free scalar field (distinct from the quantized scalar field that will be added to this background). This choice of metric has the advantage that the formalism of Sec.~\ref{subsec:quafiemod} ff.\ remains completely unchanged when comparing the inflationary Kasner and FLRW calculations (in particular, ${a_1a_2a_3/a_0=t}$).

The renormalized FLRW variance has been calculated analytically for a conformally coupled massless scalar field using both point-splitting and adiabatic techniques \cite{dav77,par74,and87,del15}. Since the spacetime is isotropic, homogeneous, and spatially flat, it is conformally Minkowski. As such, in the massless, conformally coupled case, the field can be decomposed into $\bm{k}$-modes as in Eq.~(\ref{eq:phi_modedecomp}), with mode solutions
\begin{equation}\label{eq:flrwmodes}
    \psi_{k}=\frac{\text{exp}\left(-\frac{3}{2}ikt^{2/3}\right)}{\sqrt{2kt^{2/3}}},\quad k\equiv|\bm{k}|=\sqrt{k_1^2+k_2^2+k_3^2}.
\end{equation}
These modes are positive frequency with respect to the globally timelike conformal Killing vector $\partial_\eta$ [with ${\eta\equiv\int a(t)^{-1}dt=(3/2)t^{2/3}}$]. Therefore, unlike in the case of inflationary Kasner, the modes of Eq.~(\ref{eq:flrwmodes}) uniquely define a natural vacuum state. In particular, the adiabatic vacuum state of Sec.~\ref{subsec:chovacsta} is equivalent to this vacuum state for all matching points $t_A$, to all adiabatic orders.

\begin{figure}[t!]
    \includegraphics[width=0.95\columnwidth]{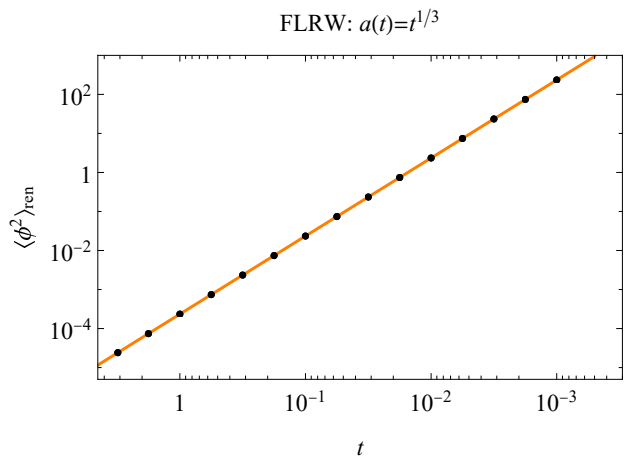}
    \caption{The renormalized variance in an FLRW background with scale factor power law index $\sfrac{1}{3}$. The black points are computed numerically using the adiabatic regularization scheme outlined in Sec.~\ref{sec:renpro}, and the orange line indicates the analytic expression from Eq.~(\ref{eq:phisq_ren_FLRW}) derived from both adiabatic and point-splitting regularization techniques.}
  \label{fig:phisq_ren_FLRW}
\end{figure}

On the numerical side, the corresponding initial conditions that must replace Eq.~(\ref{eq:inits}) to ensure the same conformal vacuum state as in the analytic case are those tied to the modes
\begin{equation}
    W_k=kt^{-1/3}.
\end{equation}
Otherwise, after the replacement of the scale factors of Eq.~(\ref{eq:flrw}) and the inclusion of the now nonzero Ricci scalar ${R=-2/(3t^2)}$, the numerical scheme presented in the previous sections can be followed exactly as in the inflationary Kasner case.

In the conformal vacuum state, the renormalized variance of a massless, conformally-coupled field over the FLRW background is \cite{bir82}
\begin{equation}\label{eq:phisq_ren_FLRW}
    \langle\phi^2\rangle_{\text{ren}}=-\frac{R}{288\pi^2}=\frac{1}{432\pi^2t^2}.
\end{equation}

The numerical adiabatic computation of ${\langle\phi^2\rangle_{\text{ren}}}$ is presented in Fig.~\ref{fig:phisq_ren_FLRW} alongside the analytic expression from Eq.~(\ref{eq:phisq_ren_FLRW}). The integrations necessary to compute each point in this figure converged to a steady value rather quickly, usually requiring an ultraviolet cutoff of no more than ${\Lambda=1}$. As shown, the two methods show excellent agreement, lending credence to the validity and precision of the present adiabatic numerical scheme.

\subsection{\label{subsec:infkasren}Inflationary Kasner renormalization}

For the inflationary Kasner spacetime, the wave Eq.~(\ref{eq:Wk}) subject to the boundary conditions of Eq.~(\ref{eq:inits}) is solved numerically using \textsc{Mathematica}'s parametric ODE solver. The mode solutions are then used to compute the integral of Eq.~(\ref{eq:phisq_ren}) for successively larger values of the momentum cutoff parameter $\Lambda$ until convergence is achieved. The result is the numerical value of ${\langle\phi^2(t)\rangle_\text{ren}}$ at a particular time $t$; this process is then repeated for different values of $t$ until the variance's full time dependence is found.

\begin{figure*}[!hb]
\centering
\begin{minipage}[l]{0.95\columnwidth}
\begin{minipage}{0.75\columnwidth}
  \centering
  \includegraphics[width=\linewidth]{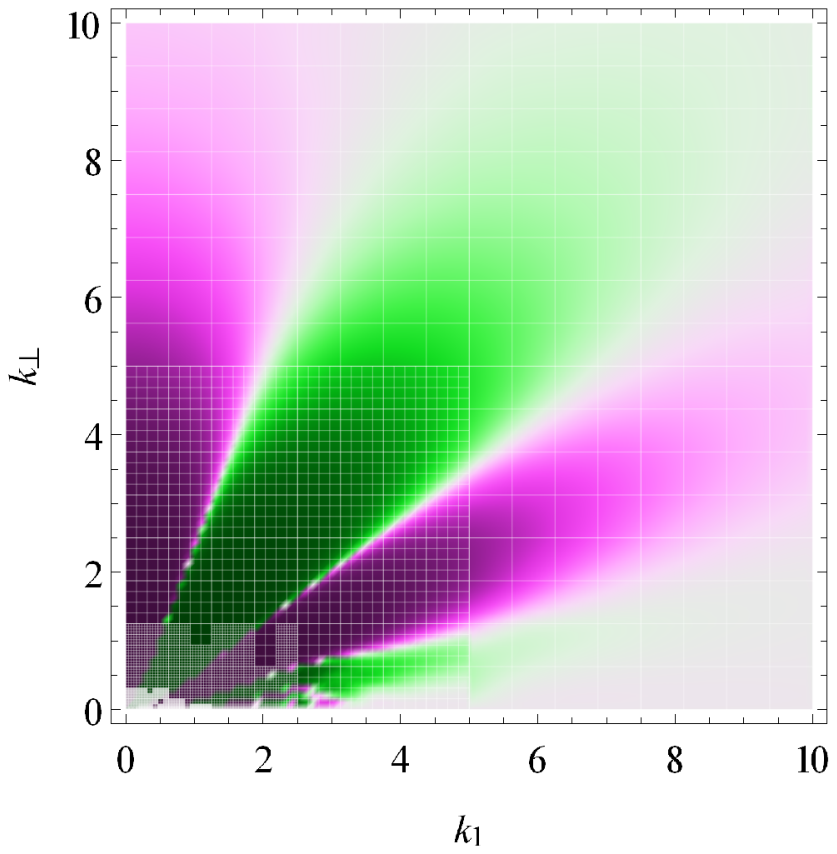}
\end{minipage}%
\begin{minipage}{.25\columnwidth}
  \centering
  \includegraphics[width=0.75\linewidth]{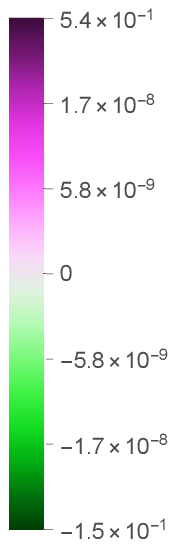}
\end{minipage}
\caption{Power spectrum [the integrand of Eq.~(\ref{eq:phisq_ren})] over a portion of $\bm{k}$-space, evaluated at time ${t=t_0}$ (as defined in Sec.~\ref{subsec:spamet}). Gridlines reveal the steps of adaptive mesh refinement, with an error threshhold of $10^{-8}$ and grid sizes varying from 0.625 to ${2\times10^{-5}}$. The constants used are the black hole spin ${a=0.96}$, initial accretion rate ${u=0.02}$, and adiabatic matching time ${t_A=5}$.\label{fig:AMR}}
\end{minipage}%
\hspace{1.5em}
\begin{minipage}[r]{\columnwidth}
    \includegraphics[width=\columnwidth]{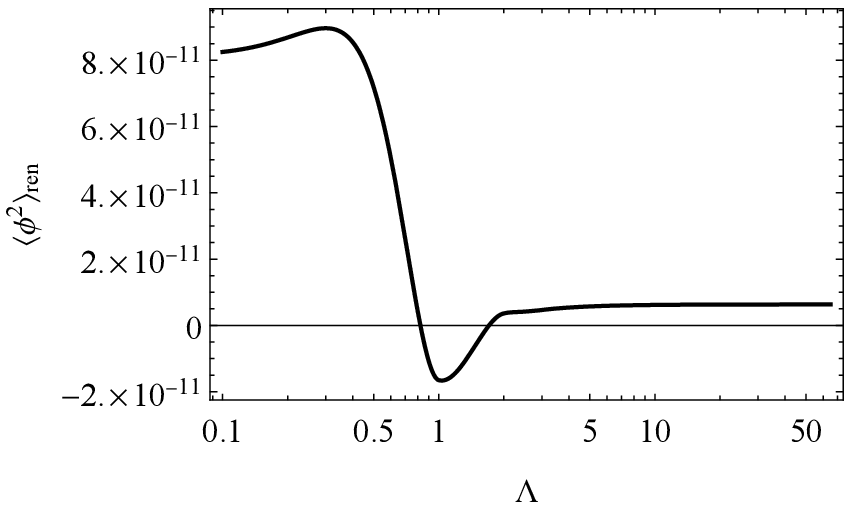}
    \caption{Convergence of the integral in Eq.~(\ref{eq:phisq_ren}) at time ${t=t_0}$ as a function of the cutoff parameter $\Lambda$. The choice of constants is the same as in Fig.~\ref{fig:AMR}.\label{fig:convergence}}
\end{minipage}
\begin{minipage}[l]{0.95\columnwidth}
  \vspace{7em}
  \includegraphics[width=\linewidth]{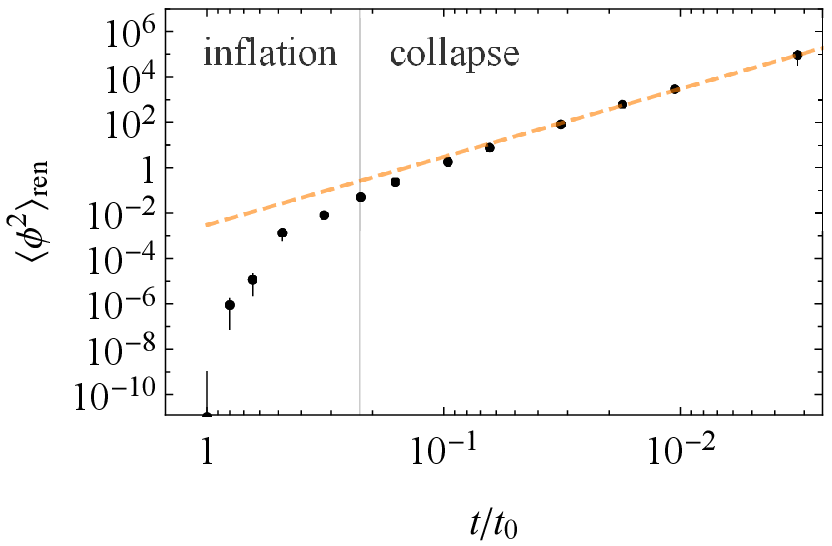}
    \caption{Renormalized adiabatic vacuum expectation value of the quantized field variance as a function of time in the inflationary Kasner spacetime. The vertical line signals the bounce transition from the mass inflation epoch to the spacelike collapse epoch, and the dashed line shows a $t^{-3}$ power law. The black hole spin is ${a=0.96}$, the initial accretion rate is ${u=0.02}$, and the adiabatic matching time is ${t_A=5}$.\label{fig:phisq_ren}}
\end{minipage}%
\hspace{1.5em}
\begin{minipage}[r]{\columnwidth}
    \includegraphics[width=0.95\columnwidth]{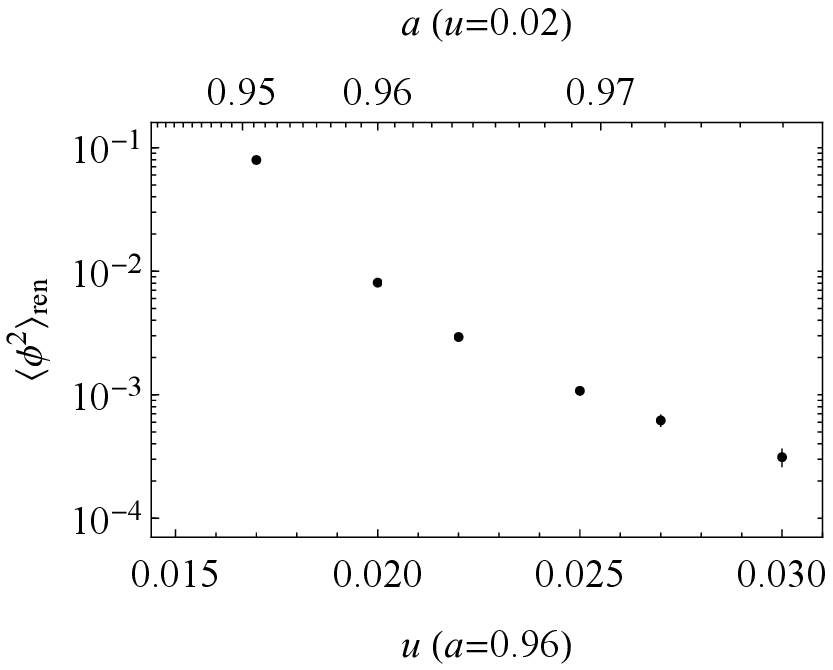}
    \caption{Renormalized adiabatic vacuum expectation value of the quantized field variance as a function of the inflationary Kasner metric coefficient $c_1$, which depends on both the initial accretion rate $u$ and the black hole spin $a$. All the values shown are calculated at a time ${t/t_0\approx0.32}$.\label{fig:phisq_ren_c1}}
\end{minipage}
\end{figure*}

Especially for small values of the inflationary Kasner time $t$, each numerical calculation of the wave equation to produce a parametric mode solution can take on the order of seconds or even minutes. As such, interpolation is used to increase the code's efficiency. A grid of points over the pertinent $\bm{k}$-space is sampled to calculate the mode solutions, and the remaining $\bm{k}$-space is estimated using third-order Hermite interpolation. To ensure that no features in the $\bm{k}$-space are overlooked by the choice of sampling grid, adaptive mesh refinement techniques are employed, such that if the errors in the interpolation function for a given region of $\bm{k}$-space are greater than a pre-determined threshold, the grid is refined to include more sampled points within that region. This process is then repeated until the integrals within all regions lie below the error threshold. An example of this process is shown in Fig.~\ref{fig:AMR}.

When the power spectrum of Fig.~\ref{fig:AMR} is integrated over both $k_1$ and $k_{\perp}$, the result is ${\langle\phi^2\rangle_\text{ren}}$. As $k_1$ or $k_{\perp}$ increases, the value of the integrand of Eq.~(\ref{eq:phisq_ren}) decreases until it reaches zero, since the numerically computed mode solutions $W_{\bm{k}}$ by construction will approach the same value as the adiabatic mode solutions of Eq.~(\ref{eq:Wk(2)}) in the large-$\bm{k}$ adiabatic limit. To ensure that enough of the infinite $\bm{k}$-space is being integrated over to calculate ${\langle\phi^2\rangle_\text{ren}}$, the integral is performed for successively larger values of $\Lambda$ until ${\langle\phi^2\rangle_\text{ren}}$ converges. As an example, Fig.~\ref{fig:convergence} shows the convergence of ${\langle\phi^2(t)\rangle_\text{ren}}$ for the same time used in Fig.~\ref{fig:AMR} (${t=t_0}$). That is, the integral of the region shown in Fig.~\ref{fig:AMR} corresponds to the point at ${\Lambda=10}$ in Fig.~\ref{fig:convergence}.

Since the time $t$ in Figs.~\ref{fig:AMR} and \ref{fig:convergence} is close to the adiabatic vacuum time $t_A$, only the lowest $\bm{k}$-modes are occupied, and the integral converges quickly. But as $t$ advances from $t_0$ to 0 and the spacetime evolves through the inflationary Kasner bounce, higher modes are expected to be occupied as the strong gravitational field seeds further particle production. Throughout the first Kasner epoch from ${t\sim t_0}$ to ${t\sim\sqrt{\sfrac{1}{2}}}$ signaling mass inflation, the transverse modes ($k_{\perp}$) become progressively more occupied, and during the second Kasner epoch from ${t\sim\sqrt{\sfrac{1}{2}}}$ to ${t\sim0}$ signaling spacelike collapse, energy from the transverse modes passes over into the higher radial modes ($k_1$).

The time evolution of ${\langle\phi^2\rangle_\text{ren}}$ for a massless scalar field in the adiabatic vacuum in the inflationary Kasner spacetime is shown in Fig.~\ref{fig:phisq_ren}. Since $t$ begins close to the adiabatic vacuum time $t_A$ ($t_0$ is about 3.1 for the choice of constants used here, while ${t_A=5}$), the vacuum polarization ${\langle\phi^2(t_0)\rangle_\text{ren}}$ begins very close to 0 (the precision-limited calculation done here for ${t=t_0}$ has error bars crossing through 0). But as mass inflation progresses, ${\langle\phi^2\rangle_\text{ren}}$ increases as more and more quantized field modes become occupied by the changing gravitational potential.

After the inflationary Kasner spacetime undergoes a bounce and proceeds to collapse towards the strong, spacelike singularity, the renormalized variance continues to increase, spanning several orders of magnitude as it approaches a divergence at ${t=0}$. Throughout the duration of the collapse epoch, ${\langle\phi^2\rangle_\text{ren}}$ seems to follow a quasi-power law in time, with an index of $-3$. Qualitatively, the vacuum polarization follows a similar trend to that of the inflationary Kasner spacetime's classical stress-energy tensor, whose density and radial flux components in a locally orthonormal tetrad frame are \cite{mcm21}
\begin{equation}
    T_{00}=T_{11}=\frac{1}{4\pi c_1t\text{ e}^{t^2}},
\end{equation}
which also asymptote to a power law in $t$ during the final collapse regime. The main difference is that the power law divergence in ${\langle\phi^2\rangle_\text{ren}}$ is even steeper than that of the classical stress-energy tensor.

\section{\label{sec:dis}Discussion}

The results from Sec.~\ref{subsec:infkasren} suggest that as an observer falls into a rotating, accreting black hole and approaches the inner horizon, the classical picture of mass inflation and subsequent spacelike collapse is reinforced when semiclassical effects are taken into account. In particular, when a quantized, massless, neutral, scalar field in the adiabatic vacuum state is coupled to the inflationary Kasner spacetime, to first-loop order, that field acquires a nonzero variance that follows a similar trend to that of the classical stress-energy of the spacetime, asymptotically approaching a power law divergence.

Since the vacuum polarization ${\langle\phi^2\rangle_\text{ren}}$ can be seen as a tracer for the behavior of the renormalized stress-energy tensor ${\langle T_{\mu\nu}\rangle_\text{ren}}$, which feeds back into the geometry of the spacetime via the semiclassical Einstein equations [see Eq.~(\ref{eq:semiclassicalEinstein})], the picture that emerges is a quantum backreaction that acts to amplify the strength of the curvature singularity at the inner horizon. Locally, particle production occurs near the inner horizon as the vacuum interacts with the inflating and collapsing spacetime curvature, and these particles seed further accretion that should feed back into the same classical inflationary Kasner spacetime.

The choice of constants used throughout this study is made both for numerical convenience and astrophysical relevance. As mentioned in Sec.~\ref{subsec:chovacsta}, the choice of the adiabatic matching time $t_A$ should not change the results of the renormalization much, as long as that time lies within the adiabatic regime. The only other independent constant in this model is $c_1$, the metric coefficient for the radial scale factor, which is determined by two physical constants via Eqs.~(\ref{eq:tc1}) and (\ref{eq:T0Phi0}), the black hole spin $a$ and the initial accretion rate $u$. Though the spin parameter $a$ can take on any value between 0 and 1, only a small range of values near 1 lead to numerically tractable values for $c_1$ with the present choice of coordinates (for example, when ${a=0.5}$, the presence of an exponential term in the conversion factor leads to ${c_1\sim10^{-70}}$). However, within this range, changing the spin does not change the qualitative behavior of Fig.~\ref{fig:phisq_ren} (only the overall magnitude), and many astrophysical black holes have been observed with spins consistent with what has been used for this analysis \cite{gou11,gar21}.

Regardless, the picture presented here is expected to hold for all astrophysically relevant ranges of spin and accretion; as either $a$ or $u$ increases, the overall magnitude of ${\langle\phi^2\rangle_\text{ren}}$ decreases nonlinearly while preserving its general qualitative trend in $t$. The dependence of ${\langle\phi^2\rangle_\text{ren}}$ on these parameters for a fixed time ${t=1\approx0.32t_0}$ (in the intermediate regime between inflation and collapse) is shown in Fig.~\ref{fig:phisq_ren_c1}. For a fixed initial accretion rate $u$, as the spin $a$ increases, ${\langle\phi^2\rangle_\text{ren}}$ decreases, and similarly, for a fixed spin $a$, as the initial accretion rate $u$ increases, ${\langle\phi^2\rangle_\text{ren}}$ decreases. Classically, the tinier the accretion rate, the more powerful mass inflation becomes, and here we find that the same holds true for the semiclassical back-reaction to mass inflation.

To ensure the robustness of the results presented here, we intend to engage in a deeper study of vacuum states near the inner horizon and build on this framework to calculate the renormalized quantum stress-energy tensor ${\langle T_{\mu\nu}\rangle_\text{ren}}$. Such a calculation involves derivatives of the mode functions and contains stronger divergences to be renormalized than those of ${\langle\phi^2\rangle_\text{ren}}$, but knowing ${\langle T_{\mu\nu}\rangle_\text{ren}}$ will allow for a more direct understanding of the quantum backreaction at the inner horizon of rotating, accreting black holes.

\bibliography{apsbib}

\end{document}